\def\draftversion{false} 
\def\showall{true}  

\RequirePackage{ifthen}
\ifthenelse{\equal{\draftversion}{true}}{
  \documentclass[
  aps,
  prb,
  10pt,
  galley,
  amsmath,amssymb,
  superscriptaddress,
  nofootinbib,notitlepage,
  showpacs,
  preprintnumbers
  ]{revtex4-1}
\usepackage{showlabels}
}{
  \documentclass[
  aps,
  prb,
  10pt,
  twocolumn,
  amsmath,
  amssymb,
  superscriptaddress,
  showpacs,
  preprintnumbers
  ]{revtex4-2}
}

\usepackage{MSGstyle}
\usepackage{my_comment}

\begin{document}

\title{Converging tetrahedron method calculations for the nondissipative parts of spectral functions}

\author{Minsu Ghim}
\email{minsu.ghim.physics@gmail.com}
\affiliation{Center for Correlated Electron Systems, Institute for Basic Science, Seoul 08826, Korea}
\affiliation{Department of Physics and Astronomy, Seoul National University, Seoul 08826, Korea}
\affiliation{Center for Theoretical Physics, Seoul National University, Seoul 08826, Korea}
\author{Cheol-Hwan Park}
\email{cheolhwan@snu.ac.kr}
\affiliation{Center for Correlated Electron Systems, Institute for Basic Science, Seoul 08826, Korea}
\affiliation{Department of Physics and Astronomy, Seoul National University, Seoul 08826, Korea}
\affiliation{Center for Theoretical Physics, Seoul National University, Seoul 08826, Korea}
\affiliation{Donostia International Physics Center, 20018 San Sebastián, Spain}
\affiliation{Centro de Física de Materiales, Universidad del País Vasco, EHU, 20018 San Sebastián, Spain}

\date{\today}

\begin{abstract}
Many physical quantities in solid-state physics are calculated from $k$-space summation.  For spectral functions, the frequency-dependent factor can be decomposed into the energy-conserving delta function part and the nondissipative principal value part.  A very useful scheme for this $k$-space summation is the tetrahedron method.  Tetrahedron methods have been widely used to calculate the summation of the energy-conserving delta function part such as the imaginary part of the dielectric function. On the other hand, the corresponding tetrahedron method for the nondissipative part such as the real part of the dielectric function has not been used much. In this paper, we address the technical difficulties in the tetrahedron method for the nondissipative part and present an easy-to-implement, stable method to overcome those difficulties. We demonstrate our method by calculating the static and dynamical spin Hall conductivity of platinum.
Our method can be widely applied to calculate linear static or dynamical conductivity, self-energy of an electron, and electric polarizability, to name a few.
\end{abstract}

\maketitle

\section{INTRODUCTION}
In the field of electronic structure calculations, many physical quantities of a periodic system are obtained from a Brillouin zone (BZ) integral of the following form:
\begin{equation}
\label{eq:spectral}
    \sum_{n, m\neq n}\int_{\textrm{BZ}} \frac{d^{3}k}{(2\pi)^{3}} f_{\nk}\frac{\Fnmk}{\hbar\omega-\left(\epsilon_{m{\bf k}}-\epsilon_{n{\bf k}}\right)+i\eta}\,.
\end{equation}
Here $f_{\nk}$ and $\epsilon_{n{\bf k}}$ are the Fermi-Dirac occupation factor and energy eigenvalue for an electronic state with band index $n$ and Bloch wavevector {\bf k}, respectively, $\Fnmk$, a complex function of two band indices $n$ and $m$ and Bloch wavevector {\bf k}, is a product of proper matrix elements, and $\eta=0^+$ is an infinitesimal positive number. Physical quantities such as correlation functions from linear response theory (conductivity, susceptibility, dielectric functions, etc\,.) or any integration of Green's functions in $k$ space (e.\,g.\,, self-energy) can be obtained  by calculating this integral.

The Brillouin-zone integration has been mainly performed from two different schemes: the smearing method and the tetrahedron method. The smearing method assumes that $\eta$ is small but finite. The integral can be replaced by a sum over a special set of $k$ points. Next, the tetrahedron method is the analytical integration method assuming that $\Fnmk$, $\epsilon_{m{\bf k}}$, and $\epsilon_{n{\bf k}}$ vary linearly inside a tetrahedron in $k$ space. Six tetrahedra are constructed for each parallelepiped in $k$ space whose vertices belong to the regular $k$-point grid: described in Appendix \hyperlink{appendix:a}{A}. In a single tetrahedron, the matrix elements and band energies are linearized.

The integral gives two terms, the principal value part and the delta function part, thanks to the following decomposition:
\begin{equation}
\label{eq:decomp}
\begin{split}
    \frac{1}{\hbar\omega-\left(\epsilon_{m{\bf k}}-\epsilon_{n{\bf k}}\right)+i\eta}&\,=\\
   {\rm P}\,\frac{1}{\hbar\omega-\left(\epsilon_{\mk}-\epsilon_{\nk}\right)}& - i\pi\delta\left(\hbar\omega-(\epsilon_{\mk}-\epsilon_{\nk})\right)\,.
\end{split}
\end{equation}
While being used to calculate the energy-conserving delta function part~\cite{macdonald1979extensions}, the tetrahedron method has not been used much for the nondissipative principal value part. In principle, the nondissipative part can be obtained from the delta function part, since the two are connected by the Kramers-Kronig relation. However, the Kramers-Kronig transform may require a significant amount of computational time: The Kramers-Kronig relation is written as an integral with respect to the frequency from zero to infinity. It is thus required to calculate the delta function part at a dense frequency grid covering a wide frequency range, even if a single frequency component of the nondissipative part is needed.

Previous studies suggested the tetrahedron method for the nondissipative part~\cite{ gilat1975tetrahedron, brener1981matrix}. The method can also be applied to metals with a proper division of tetrahedra at the Fermi surface at zero temperature~\cite{lehmann1972numerical}. However, this method has some technical difficulties. First, the explicit formula for the integration is in seven different complicated forms, depending on the energy eigenvalues at the four vertices of a tetrahedron. In more recent studies~\cite{Kaprzyk_1986, KAPRZYK2012347}, the multiple-case method was similarly adopted. Second, the criteria to distinguish between these cases are not well-defined numerically: They depend on whether some variables are exactly zero or not. Third, most importantly, the round-off error occurring during the evaluation of the logarithmic functions is not controlled. To implement the tetrahedron method for the nondissipative part, these difficulties should be handled appropriately.

In this paper, we present an easy-to-implement, numerically stable tetrahedron method for the nondissipative part to overcome these technical difficulties.
After a detailed explanation of the method, we compare the convergence and the computational time of the tetrahedron method with those of the adaptive smearing technique~\cite{yates2007spectral} for the intrinsic static and dynamical spin Hall conductivity (SHC) of fcc platinum using the interpolation scheme based on maximally locallized Wannier functions (MLWFs)~\cite{souza2001maximally,ryoo2019computation}. We show that, to reach the same level of convergence, our improved tetrahedron method requires orders of magnitude shorter computation times than the adaptive smearing method. Finally, we discuss how the convergence can be achieved efficiently.

\section{METHOD}
\subsection{{\rm \bf Tetrahedron Method for the Kubo Formula}}

Consider the following three types of integrals:
\begin{align} \label{eq:tetinteg1}
    I_{1,nm}(\hbar\omega)&=\int_{\textrm{tet.}} d^{3}k\, {\rm P}\,\frac{\Fnmk}{\Dnmk+\hbar\omega}\,,&\\
    \label{eq:tetinteg2}
    I_{2,nm}(\hbar\omega)&=\int_{\textrm{tet.,} \Dnmk=\hbar\omega} d^{2}S_{k}\frac{\Fnmk}{|\nabla_{\textbf{k}}\Dnmk|}\,, &\\
    \label{eq:tetinteg3}
    I_{3,nm}&=\int_{\textrm{tet.}} d^{3}k \frac{\Fnmk}{\Dnmk^{2}}\,,
\end{align}
where the integration in momentum space is performed over a particular tetrahedron, and $\Dnmk=\epsilon_{\nk} - \epsilon_{\mk}$. Especially, $I_{1,nm}$ and $I_{3,nm}$ is obtained by evaluating the nondissipative parts.  The formulas to calculate Green's-function-like or susceptibility-like quantities such as anomalous hall conductivity~\cite{PhysRevB.53.3692, RevModPhys.82.1539}, Fan-Migdal self-energy~\cite{RevModPhys.89.015003, PhysRev.82.900,migdal1958interaction}, and spin Hall conductivity~\cite{guo2005ab, guo2008intrinsic}, can be transformed and decomposed to terms of the forms in Eqs.~\eqref{eq:tetinteg1}\,–\,\eqref{eq:tetinteg3}.  To allow analytical integration in Eqs.~\eqref{eq:tetinteg1}, \eqref{eq:tetinteg2}, and~\eqref{eq:tetinteg3}, we used the Pad\'e approximant, the best approximation by a rational function. The Pad\'e approximant is obtained by using \begin{align}
\label{eq:linear_interpolation}
    \Fnmk =&\, F_0 + F_x k_x + F_y k_y + F_z k_z\\
\label{eq:linear_interpolation2}
    \Delta_{nm\mathbf{k}} =&\, \Delta_0 + \Delta_x k_x + \Delta_y k_y + \Delta_z k_z
\end{align}
where $F_0, \,\cdots, F_z$, and $\Delta_0, \,\cdots, \Delta_z$ are the parameters reproducing the values at each of the four vertices of a tetrahedron. As a result,
\begin{align} \label{eq:tetinteg1_linearized}
    I_{1,nm}(\hbar\omega)&=\int_{\textrm{tet.}} d^{3}k\, {\rm P}\,\frac{F_0 + F_x k_x + F_y k_y + F_z k_z}{\Delta_0 +\hbar\omega + \Delta_x k_x + \Delta_y k_y + \Delta_z k_z}&\\
    \label{eq:tetinteg3_linearized}
    I_{3,nm}&=\int_{\textrm{tet.}} d^{3}k \frac{F_0 + F_x k_x + F_y k_y + F_z k_z}{[\Delta_0 + \Delta_x k_x + \Delta_y k_y + \Delta_z k_z]^{2}}\,,
\end{align}
and these can be integrated by hand since we know that the integration of rational functions contains other rational functions with logarithms. For example, the one-dimensional case for Eq.~\eqref{eq:tetinteg1_linearized} is equivalent to
\begin{align}
\begin{split}
    \int_{k_i}^{k_f} dk_x &{\rm P}\,\frac{F_0 + F_x k_x}{\Delta_0 + \hbar\omega + \Delta_x k_x}\\
    =&\, \frac{F_0 \Delta_x - F_x(\Delta_0+\hbar\omega)}{\Delta_x^2}{\rm log}\left|\frac{\Delta_x k_f + \Delta_0 + \hbar\omega}{\Delta_x k_i + \Delta_0 + \hbar\omega}\right|\\
    &+ \frac{F_x}{\Delta_x}(k_f-k_i)\,.
\end{split}
\end{align}
The closed-form expressions for the three-dimensional case are presented in Appendix~\hyperlink{appendix:b}{B}. We note that the equivalent results for $I_{1,nm}$ [Eq.~\eqref{eq:tetinteg1}] are in Ref.~\cite{brener1981matrix} and those for $I_{2,nm}$ [Eq.~\eqref{eq:tetinteg2}] are in Ref.~\cite{macdonald1979extensions}. The results for $I_{3,nm}$ [Eq.~\eqref{eq:tetinteg3}] are presented here for the first time.

A specific component of the intrinsic spin Hall conductivity, for example, is expressed as~\cite{guo2005ab, guo2008intrinsic}
\begin{align} \label{eq:Kubo0}
\begin{split}
    \sigma^{z}_{xy}(\omega)={}\frac{e}{\hbar}&\frac{\Omega}{(2\pi)^{3}}\sum_{n, m\neq n}\int_{\textrm{BZ}} d^{3}k (f_{\nk}-f_{\mk})\times\\
    &\frac{{\rm Im}[\mel{u_{\nk}}{\hat{j}^{z}_{x\textbf{k}}}{u_{\mk}} \mel{u_{\mk}}{\hat{v}_{y\textbf{k}}}{u_{\nk}}]}{\Dnmk^{2}-(\hbar\omega+i\eta)^{2}}
\end{split}
\end{align}
where $\Omega$ is the volume of a unit cell and $\ket{u_{\nk}}$ is the periodic part of the Bloch state satisfying $H_{\mathbf{k}}\ket{u_{\nk}}=\epsilon_{\nk}\ket{u_{\nk}}$ where $H_{\mathbf{k}}=e^{-i\mathbf{k}\cdot\mathbf{r}} H e^{i\mathbf{k}\cdot\mathbf{r}}$, and $H$ is a lattice-periodic Hamiltonian. $\hat{j}^{z}_{x\textbf{k}}=\frac{\hbar}{4}\{\sigma^{z},v_{x\textbf{k}}\}$, and $\hat{v}_{y\textbf{k}}=\frac{1}{\hbar}\partial_{y}H_{\textbf{k}}$ are the spin current and velocity operators, respectively. After dividing the Brillouin zone according to the scheme presented in Appendix \hyperlink{appendix:a}{A}, we can decompose Eq.~\eqref{eq:Kubo0} into integrations over a single tetrahedron of the forms in Eqs.~\eqref{eq:tetinteg1}--\eqref{eq:tetinteg3}.
We obtain the dynamical conductivity of the form
\begin{align}
\begin{split}
    &\frac{1}{2\hbar\omega}\sum_{n, m\neq n}\left[\left\{I_{1,nm}(-\hbar\omega)-I_{1,nm}(\hbar\omega)\right\}\right.\\&\left.+i\pi \left\{I_{2,nm}(-\hbar\omega)+I_{2,nm}(\hbar\omega)\right\}\right]
\end{split}
\end{align}
and the static conductivity of the form
\begin{align}
    \sum_{n, m\neq n}I_{3,nm}\,.
\end{align}

\subsection{{\rm \bf Numerical Problem and Its Solution}}

If we perform the integrations in Eqs.~\eqref{eq:tetinteg1} and~\eqref{eq:tetinteg3} analytically, we obtain the following results:
\begin{align} 
\begin{split} \label{eq:tetinteg1_again}
    I_{1, nm}(\hbar\omega) ={}& -\frac{(1+x_1)(1+x_2)(1+x_3)}{6x_1^2 x_2^2 x_3^2 (x_1 - x_2)^2 (x_2 - x_3)^2 (x_3 - x_1)^2}\\
    & \times \frac{\textrm{det}(\textbf{t})}{\Delta_4 + \hbar\omega} \sum_{i = 1}^{4}F_{i}\left(\sum_{j=1}^{3}C_{ij}^{(1)}\xi_{j} + B_{i}^{(1)}\right)
\end{split}\\
\begin{split} \label{eq:tetinteg3_again}
    I_{3, nm} ={}&  \frac{(1+x_1)(1+x_2)(1+x_3)}{2x_1^2 x_2^2 x_3^2 (x_1 - x_2)^2 (x_2 - x_3)^2 (x_3 - x_1)^2}\\
    & \times \frac{\textrm{det}(\textbf{t})}{\Delta_4^2} \sum_{i = 1}^{4}F_{i}\left(\sum_{j=1}^{3}C_{ij}^{(3)}\xi_{j} + B_{i}^{(3)}\right)\,,
\end{split}
\end{align}
where $F_{i}$ and $\Delta_{i}$ ($i\in\{1,2,3,4\}$) are the values of $\Fnmk$ and $\Dnmk$ at the four vertices of a single tetrahedron, respectively. $x_j$ and $\xi_j$ ($j\in\{1,2,3\}$) are defined as:
\begin{align}
    x_j =& \frac{\Delta_4 - \Delta_j}{\Delta_j + \hbar\omega},\,\xi_j = \textrm{log}|1 + x_j| \,.
\end{align}
The expressions for $C_{ij}$ and $B_{i}$ in terms of $x_i$'s are given in Appendix \hyperlink{appendix:b}{B}.

However, there remains a numerical problem of calculating a very small number divided by another very small number with floating-point arithmetic, which occurs when the values of $\Dnmk$ of two or more vertices are very close to each other.
According to the definition above, $x_1$, $x_2$, $x_3$, $x_1-x_2$, $x_2-x_3$, and $x_3-x_1$ are measures of how close $\Delta_{i}$ and $\Delta_{j}$ are for pairs $(i, j)$ = $(1, 4), (2, 4), (3, 4), (1, 2), (2, 3)$, and $(3, 1)$, respectively. As two or more $\Delta_{i}$'s get closer, the elements in a subset $X \subseteq \{x_1, x_2, x_3, x_1-x_2, x_2-x_3, x_3-x_1\}$, become closer to zero. Then, let us define $I_{1, nm}|_{X \rightarrow 0}$ and $ I_{3, nm}|_{X \rightarrow 0}$ as the limits of $I_{1, nm}$ and $I_{3, nm}$ if all the elements in $X$ approach zero. (The expressions 
for $I_{1, nm}|_{X \rightarrow 0}$ are presented in Ref.~\cite{brener1981matrix}). However, errors are generated in evaluating the logarithmic functions because of rounding of floating-point arithmetic and are amplified due to the small denominator in the prefactors of Eqs.~\eqref{eq:tetinteg1_again} and~\eqref{eq:tetinteg3_again}, resulting in non-convergence.

Let us define $\epsilon$ as the maximum value which a computer cannot store when it is added to $1$ due to rounding of floating-point arithmetic. Roughly speaking, $\epsilon$ $\approx$ $10^{-8}$, $10^{-16}$, and $10^{-32}$ for single-, double-, and quadruple-precision data types, respectively. Explicitly, we may write this as $1 \doteq 1 + O(\epsilon)$, where $\doteq$ means ``equality in machines." For example, if a machine cannot distinguish two real numbers $A$ and $B$, $A \doteq B$. For a real number $x$, $x \doteq x + O(x\epsilon)$.

A round-off error comes from calculating $\xi_{j} = \textrm{log}|1+x_j| = O(x_j)$ (in case $|x_j| \ll 1$) or $\xi_k = \textrm{log}|1+x_k| = \textrm{log}|1+x_j+(x_k - x_j)| = \xi_j + O(x_j - x_k)$ (in case $|x_j - x_k| \ll |1 + x_j|$). Since the default computer algorithm to compute ${\rm log}$ uses the Taylor expansion, ${\rm log}\,(1+x)$ is calculated by $(1+x-1) - (1+x-1)^2/2 + (1+x-1)^3/3 + \cdots$, not by $x - x^2/2 + x^3/3 + \cdots$. The value of $1+x-1$ is not the same as that of $x$, but $1+x-1 \doteq x + O(\epsilon)$ if $|x|\ll1$. Hence, 
\begin{equation}
    \textrm{log}|1+x| \doteq O(x) + O(\epsilon)
\end{equation}
for small $|x|$ ($\ll1$). For a similar reason, log$|1+x| = \textrm{log}|1+y+(x-y)| \doteq \textrm{log}|1+y| + O(x-y) + O(\epsilon)$ for small $|x-y|$ ($\ll1$).

This round-off error can be problematic if $x_j \rightarrow 0$ or $x_j - x_k \rightarrow 0$. In the first case, a term such as
\begin{align}
\label{eq:xj_0_error}
\begin{split}
    &\frac{\sum_l C_{il}\xi_{l}+ B_i}{x_1^2 x_2^2 x_3^2 (x_1 - x_2)^2 (x_2 - x_3)^2 (x_3 - x_1)^2}\qquad\quad\\
    &\doteq O(1)\times   \frac{\sum_{l\neq j}C_{il}\xi_{l} + C_{ij}(O(x_j)+O(\epsilon)) + B_i}{x_j^2}\\
    &= O(1)\times \frac{(\sum_{l\neq j} C_{il}\xi_{l} + C_{ij}O(x_j)+ B_i) + C_{ij}O(\epsilon) }{x_j^2}\\
    &= \frac{O(x_j^2) + O(\epsilon) }{x_j^2}
\end{split}
\end{align}
should be finite as $x_j \rightarrow 0$, but is divergent due to $O(\epsilon)$.
In the last equality, we have used the fact that the coefficients of $O(x_j)$ should vanish in order to have a non-divergent value in the limit $x_j \to 0$ if $\epsilon=0$. We can also check this equality using Eqs.~\eqref{eq:C_ij_1} and~\eqref{eq:C_ij_3} in Appendix \hyperlink{appendix:b}{B}.
In the second case,
\begin{align}
\label{eq:xj_xk_0_error}
\begin{split}
    &\frac{\sum_l C_{il}\xi_{l}+ B_i}{x_1^2 x_2^2 x_3^2 (x_1 - x_2)^2 (x_2 - x_3)^2 (x_3 - x_1)^2}\qquad\quad\\
    &\doteq O(1)\times  \Bigg[\frac{\sum_{l\neq k} C_{il}\xi_{l} + B_i}{(x_j-x_k)^2}\\
    &\qquad\qquad\quad  + \frac{C_{ik}(\xi_{j} + O(x_j-x_k)+O(\epsilon)) }{(x_j-x_k)^2}\Bigg]\\
    &= O(1)\times   \Bigg[\frac{\sum_{l\neq k} C_{il}\xi_{l} + C_{ik}(\xi_{j} + O(x_j-x_k))+ B_i}{(x_j-x_k)^2}\\
    & \qquad\qquad\quad  + \frac{ C_{ik}O(\epsilon) }{(x_j-x_k)^2}\Bigg]\\
    &= \frac{O((x_j-x_k)^2) + O(\epsilon) }{(x_j-x_k)^2}
\end{split}
\end{align}
should also be finite as $x_j-x_k \rightarrow 0$, but is divergent due to $O(\epsilon)$. In the last equality, the coefficients of $O(x_j - x_k)$ should vanish similarly as Eq.~\eqref{eq:xj_0_error}.
The same phenomenon happens when two or more $x_j, x_j - x_k$ are close to zero.

It is also well known that $\textrm{log1P}$ is a more accurate way to compute $\textrm{log}(1+x)$ in case $|x| \ll 1$~\cite{beebe2017mathematical}, but the round-off error does not completely disappear here either. It is expressed as $\textrm{log1P}(x) = x\,\textrm{log}(1+x)\,/\,(1+x-1)$. The error is $O(x\epsilon)$, not $O(\epsilon)$, because
\begin{align}
\begin{split}
    \textrm{log1P}(x) =& \, x\,\frac{\textrm{log}(1+x)}{1+x-1} \\
    = & \, x\,\frac{(1+x-1) - (1+x-1)^2/2 + \cdots}{1+x-1} \\
    = & \, x\,\left[1 - (1+x-1)/2+ \cdots\right] \\
    \doteq & \, x\,(O(1) + O(\epsilon))\\
    = & \, O(x) + O(x\epsilon) \,.
\end{split}
\end{align}
Therefore, the numerical problem is not yet in general resolved even with log1P function because of the remaining error term.

To avoid this difficulty, we may first consider using the formulas for $I_{1, nm}|_{X \rightarrow 0}$ in Ref.~\cite{brener1981matrix} and finding and using the formulas for $I_{3, nm}|_{X \rightarrow 0}$. However, this method has two problems in practice. First, since there are six parameters in total, many different limiting cases should be considered depending on which of the six parameters are much smaller in magnitude than one, making the implementation complicated. Second, since they are real numbers which are not exactly zero, it is not easy to define the criteria to determine whether one of the limiting-case formulas should be applied or not.

By introducing a criterion for determining whether the elements of $X$ are close to zero, we can solve the numerical problem in a simple way, without implementing many cases. Let us redefine $x_1$, $x_2$, $x_3$, $x_1-x_2$, $x_2-x_3$, and $x_3-x_1$ symmetrically:
\begin{equation}
    x_{ij} = \frac{\Delta_{i}-\Delta_{j}}{(\Delta_{i}+\Delta_{j})/2+\hbar\omega}\,.
    \label{eq:x_ij}
\end{equation}
If one of $x_{ij}$ gets close to zero, the numerical integration will have the error shown in Eqs.~\eqref{eq:xj_0_error} and~\eqref{eq:xj_xk_0_error}. In order to prevent this error, we enforce $|x_{ij}|$ to be not smaller than a specified value $\cutoff$ (cutoff) by modifying $\Delta_{i}$ and $\Delta_{j}$ as
\begin{equation}
\label{eq:redefinition_upper}
    \Delta_{i} \rightarrow \frac{\Delta_{i}+\Delta_{j}}{2} \pm 0.5\left|\frac{\Delta_{i}+\Delta_{j}}{2}+\hbar\omega\right| \cutoff
\end{equation}
and
\begin{equation}
\label{eq:redefinition_lower}
    \Delta_{j} \rightarrow \frac{\Delta_{i}+\Delta_{j}}{2} \mp 0.5\left|\frac{\Delta_{i}+\Delta_{j}}{2}+\hbar\omega\right| \cutoff \,,
\end{equation}
where the upper signs are selected if $\Delta_{i} \ge \Delta_{j}$ and the lower signs are selected if $\Delta_{i} < \Delta_{j}$.
Accordingly, the updated $|x_{ij}|=\cutoff$, and the round-off error is not amplified by the small denominator [see Eqs.~\eqref{eq:xj_0_error} and~\eqref{eq:xj_xk_0_error}].

The cutoff should make the error term in the denominator $O(\epsilon)$ small enough compared to the true value term, $O(x_j^2)$ in Eq.~\eqref{eq:xj_0_error} or $O((x_j - x_k)^2)$ in Eq.~\eqref{eq:xj_xk_0_error}. $\cutoff$ prevents the error from becoming comparable to the true value with the condition,
\begin{equation}
    O(\epsilon) \ll O(\cutoff^2)\,.
\end{equation}
Therefore, $\cutoff$ depends on the precision, or, equivalently, the data type. If a higher-precision data type is used, the required $\cutoff$ is smaller. We adopted $\cutoff$ = $10^{-3}$ for double precision numbers, and we suggest using this value for other materials. Since the material-dependent variables, $F_i$ (velocity and spin-current matrix elements) and $x_{ij}$ (determined by band structure), are not included in the condition above, the same value would work well for other materials.

\subsection{{\rm \bf Higher-Order Correction}}
In the ordinary tetrahedron method, the matrix elements and the energy eigenvalues are linearly interpolated using the values at the four vertices. The integrand can thus deviate from the true value due to the neglected curvature in $k$ space of the numerator and the denominator. To correct this error, Blöchl suggested an improved tetrahedron method~\cite{blochl1994improved}, which has been widely implemented in many computer programs. The method can take the curvature effect into account using only the matrix elements and energy eigenvalues at the four vertices without higher-order interpolation using the Gauss theorem. Quantities such as the total energy or the charge (spin) density can be calculated using the Blöchl's correction. However, this method is limited to smoothly varying functions which are interpolated by polynomials, not by rational functions. Near the zero of the denominator in $k$ space where $\Dnmk^2 - (\hbar \omega + i\eta)^2 = 0$, the integrand is rapidly varying. The gradient and the curvature inside a tetrahedron are not well represented, resulting in the failure of the method.

Kawamura {\it et al.} introduced a novel, improved tetrahedron method applicable to response functions~\cite{kawamura2014improved}. The work is about finding a good linear interpolation function from the third-order correction. The method uses additional 16 $k$ points surrounding the original tetrahedron together with the four original vertices, and finds the 20 coefficients to construct third-order interpolating polynomials for the matrix element and for the energy eigenvalue. Finally, one fits a linear function using the least square method. As a result, the formulas for the linear fitting functions are automatically calculated from the matrix elements and the energy eigenvalues at the 20 $k$ points. This scheme can also be employed with our solution to the numerical problem (round-off errors) at the same time.

\section{COMPUTATIONAL DETAILS}

The intrinsic contribution to both the static and dynamical SHC of platinum was investigated by the simple summation with the adaptive smearing scheme~\cite{yates2007spectral} and the tetrahedron method proposed here. The entire procedure from the self-consistent calculations to the Wannier-interpolation is the same as that in Ref.~\cite{ryoo2019computation}. The norm-conserving pseudopotential and the \pbesol exchange-correlation energy functional~\cite{perdew2008restoring} were used. The kinetic energy cutoff for the wave functions was set to 60 Ry. The energy eigenvalues, the corresponding Fermi-Dirac occupancy factor, and the matrix elements for the SHC were computed via self-consistent and non-self-consistent calculations on  $12 \times 12 \times 12$ and $8 \times 8 \times 8$ $k$-point grids, respectively, using the plane-wave pseudopotential code PWSCF from the \qe  package~\cite{giannozzi2009quantum,giannozzi2017advanced}, via the \wnt  package, and via the \pwtwnt which is the connecting program between PWSCF and \wnt~\cite{pizzi2020wannier90}. The initial guess at MLWFs were $s$, $p$, and $d$ atom-centered nodeless orbitals, and the inner energy window is 15 eV's wide from the bottom of the valence-band minimum up to 4 eV above the Fermi level.

Regarding the adaptive smearing for the simple summation, we followed the scheme of Ref.~\cite{yates2007spectral}: We set $\eta$ in Eq. \eqref{eq:Kubo0} as $\eta=a|\frac{\partial \epsilon_{\nk}}{\partial \textbf{k}} - \frac{\partial \epsilon_{\mk}}{\partial \textbf{k}}| \Delta k$ where $a$ is a dimensionless constant of order one (we set $a=\sqrt{2}$), and $\Delta k$ is the distance between the nearest-neighbor $k$ points in the interpolation grid. To avoid the aforementioned numerical instability for the tetrahedron method, we set $\cutoff$ to be $10^{-3}$.

\section{RESULTS}
\begin{figure}
    \includegraphics[trim={0 0 0 0.4cm},clip,width=0.9\columnwidth]{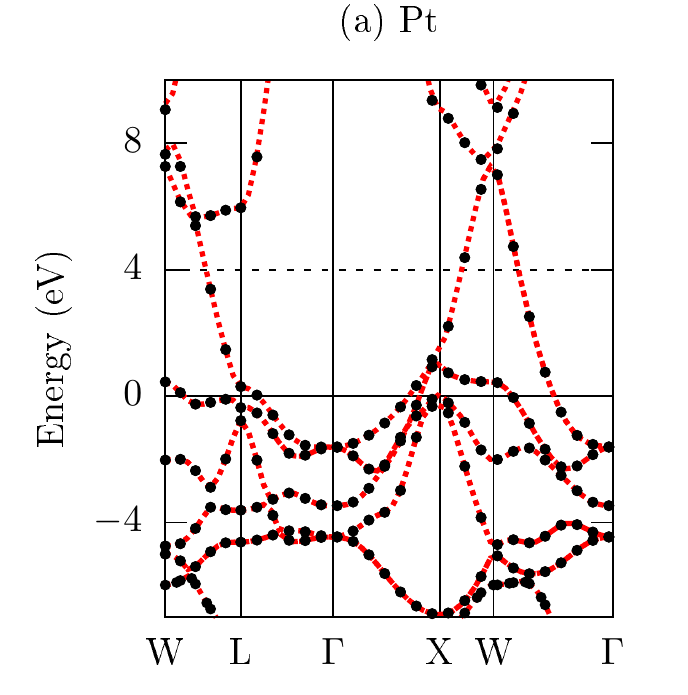}
    \caption{The combined electronic band structure of fcc platinum. The black dots indicate the band structure obtained from \textit{ab initio} calculations, and the red curves were obtained from the Wannier interpolation. The Fermi level of undoped platinum is located on the top of the valence band, as displayed by a solid horizontal line. The black dashed horizontal line is the ceiling of the inner window.}
    \label{fig:wannierization}
\end{figure}

\begin{figure*}
    \includegraphics[trim={0 3.3cm 0 1.3cm},clip, width=0.8\textwidth]{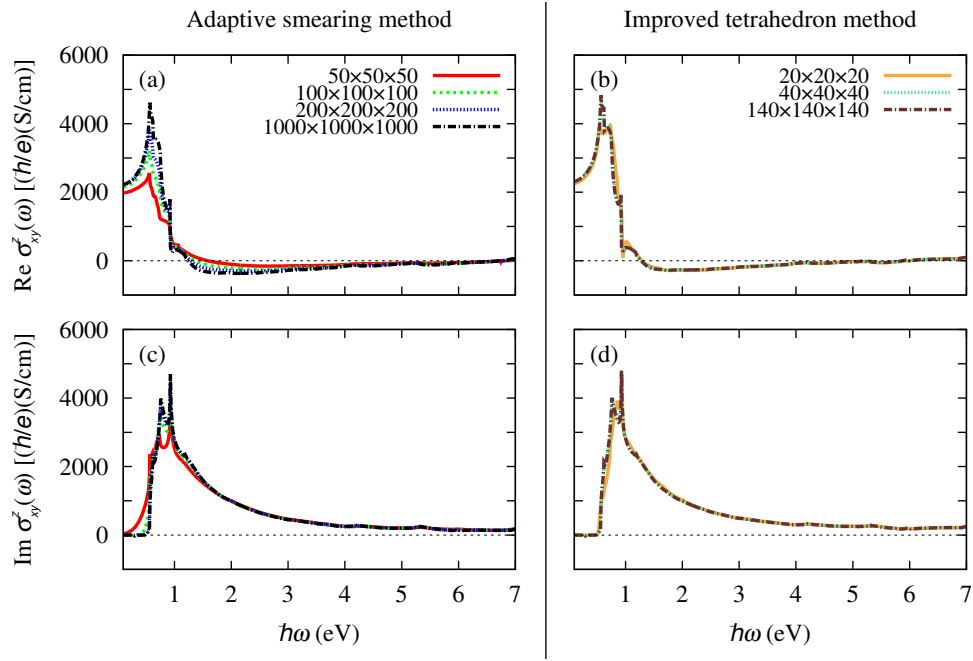}
    \caption{The dynamical SHC of platinum calculated by the simple summation with the adaptive smearing method and by the tetrahedron method with higher-order correction. The $k$-point grid used for the numerical integration is specified.}
    \label{fig:convergence}
\end{figure*}

The band structure of platinum is shown in Fig.~\ref{fig:wannierization}. 18 MLWFs successfully reproduced the band structure within the inner window ($<4$~eV).

Figure ~\ref{fig:convergence} shows the comparison of the two methods for the dynamical SHC. We used unshifted Monkhorst-Pack grids. 
The adaptive smearing scheme requires a finer $k$ mesh to reach convergence. The tetrahedron method achieved convergence at a relatively coarse $40\cross40\cross40$ grid. On the other hand, the peak value at 0.57 eV of the real part obtained on $100\cross100\cross100$ grids, using the adaptive smearing, does not reach half of that obtained on $40\cross40\cross40$ grids using the tetrahedron method.

The imaginary part is zero below 0.48~eV, which is the threshold for a direct band-to-band transition. The tetrahedron method is useful to see this feature, since it reproduces the delta function better than the smearing method. This result agrees with previous studies on the original tetrahedron method for the integrals over the Fermi surface, such as in the case of density of states.

\begin{figure*}
    \includegraphics[trim={0 3.3cm 0 1.3cm}, clip, width=1.0\textwidth]{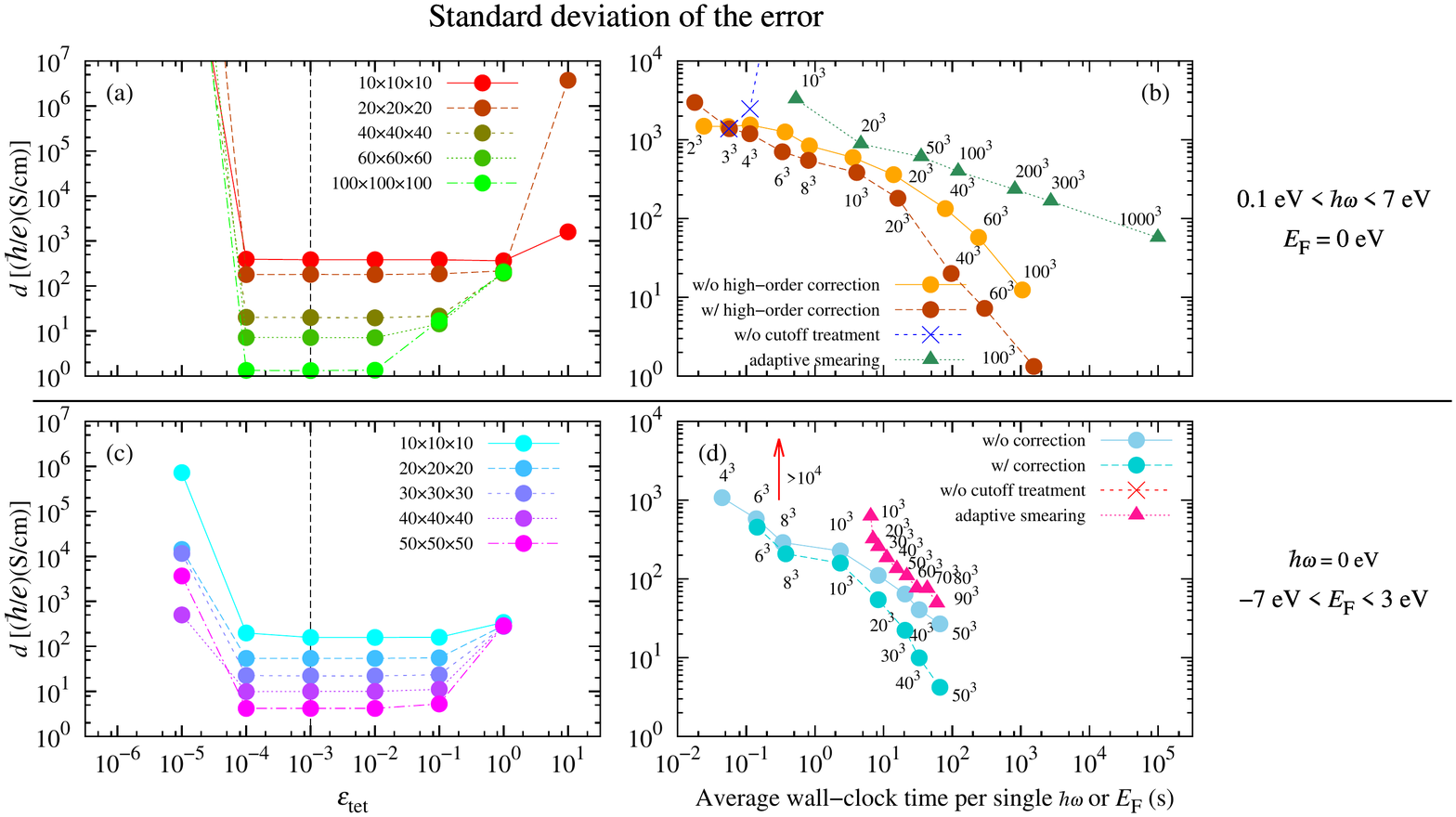}
    \caption{(a) $d_{\rm dyn.}$ [Eq.~\eqref{eq:def_of_d}] as a function of $\cutoff$. (b) A comparison of computational time among the three tetrahedron methods and the adaptive smearing method. (c) and (d) Similar quantities as (a) and (b), respectively, for $d_{\rm stat.}$ [Eq.~\eqref{eq:def_of_d_static}]. The vertical dashed line in (a) and (c) denotes $\cutoff = 10^{-3}$ selected for other calculations [the results shown in Fig.~\ref{fig:convergence} and Figs.~\ref{fig:convergence2}(b) and~\ref{fig:convergence2}(d)]. }
    \label{fig:convergence2}
\end{figure*}

To quantify convergence, we defined the standard deviation of the dynamical SHC as follows,
\begin{equation}
    d_{\rm dyn.} = \sqrt{\ddfrac{\int_{\omega_{1}}^{\omega_{2}} d\omega |\sigma^{z}_{xy}(\omega) - \sigma^{z}_{xy, {\rm conv.}}(\omega)|^2}{\omega_{2}-\omega_{1}}}\,,
    \label{eq:def_of_d}
\end{equation}
where $\omega$ is the frequency, $\omega_{1} = 0$~eV, and $\omega_{2} = 7$~eV. For the static SHC, the Fermi-energy shift $E_{\rm F}$ is substituted for $\omega$,
\begin{equation}
    d_{\rm stat.} = \sqrt{\ddfrac{\int_{E_{{\rm F}1}}^{E_{{\rm F}2}} dE_{\rm F} |\sigma^{z}_{xy}(0) - \sigma^{z}_{xy, {\rm conv.}}(0)|^2}{E_{{\rm F}2}-E_{{\rm F}1}}}\,,
    \label{eq:def_of_d_static}
\end{equation}
where $E_{{\rm F}1} = -7.1$~eV and $E_{{\rm F}2} = +3$~eV. The integrals were evaluated by a summation using $\Delta \omega = 0.1$~eV and $\Delta E_{\rm F} = 0.1$~eV. Using the tetrahedron method, the SHC calculated on $140 \times 140 \times 140$ grids for the dynamical and static SHC was considered the converged one ($\sigma^{z}_{xy, {\rm conv.}}$).

To see the cutoff dependence of our tetrahedron method, we obtained $d_{\rm dyn.}$ and $d_{\rm stat.}$ with five different $k$ meshes as a function of $\cutoff$ values (Fig.~\ref{fig:convergence2}). In general, the smaller $\cutoff$ is, the larger the round-off error is. On the other hand, the larger $\cutoff$ is, the larger the deviation of the energy eigenvalues at the four vertices of a tetrahedron [Eqs.~\eqref{eq:redefinition_upper} and~\eqref{eq:redefinition_lower}] is. Therefore, as Figs.~\ref{fig:convergence2}(a) and~\ref{fig:convergence2}(c) show, there is an optimal range for $\cutoff$ to converge SHC, with which the standard deviation of the error ($d_{\rm stat.}$ or $d_{\rm dyn.}$) is small enough compared to the total SHC value $\sigma^{z}_{xy} \sim O(10^{3}) (\hbar/e) {\rm (S/cm)}$. The dashed vertical lines indicate $\cutoff = 10^{-3}$ located in the middle of this optimal range. This value was adopted in obtaining the results shown in Fig.~\ref{fig:convergence} and Figs.~\ref{fig:convergence2}(b) and~\ref{fig:convergence2}(d).

We compared the computational time to see which method is the most effective. Figures~\ref{fig:convergence2}(b) and~\ref{fig:convergence2}(d) show, respectively, that $d_{\rm dyn.}$ and $d_{\rm stat.}$ decrease as the $k$-point grid for integration becomes dense. Convergence is not reached at all if tetrahedron method without the cutoff treatment is used, whereas the tetrahedron method with the cutoff treatment achieves convergence faster than the adaptive smearing method. The higher-order correction proposed in Ref.~\cite{kawamura2014improved} improves the convergence further.

\begin{figure}
    \includegraphics[trim={2cm 0cm 2cm 0cm}, width=0.9\columnwidth]{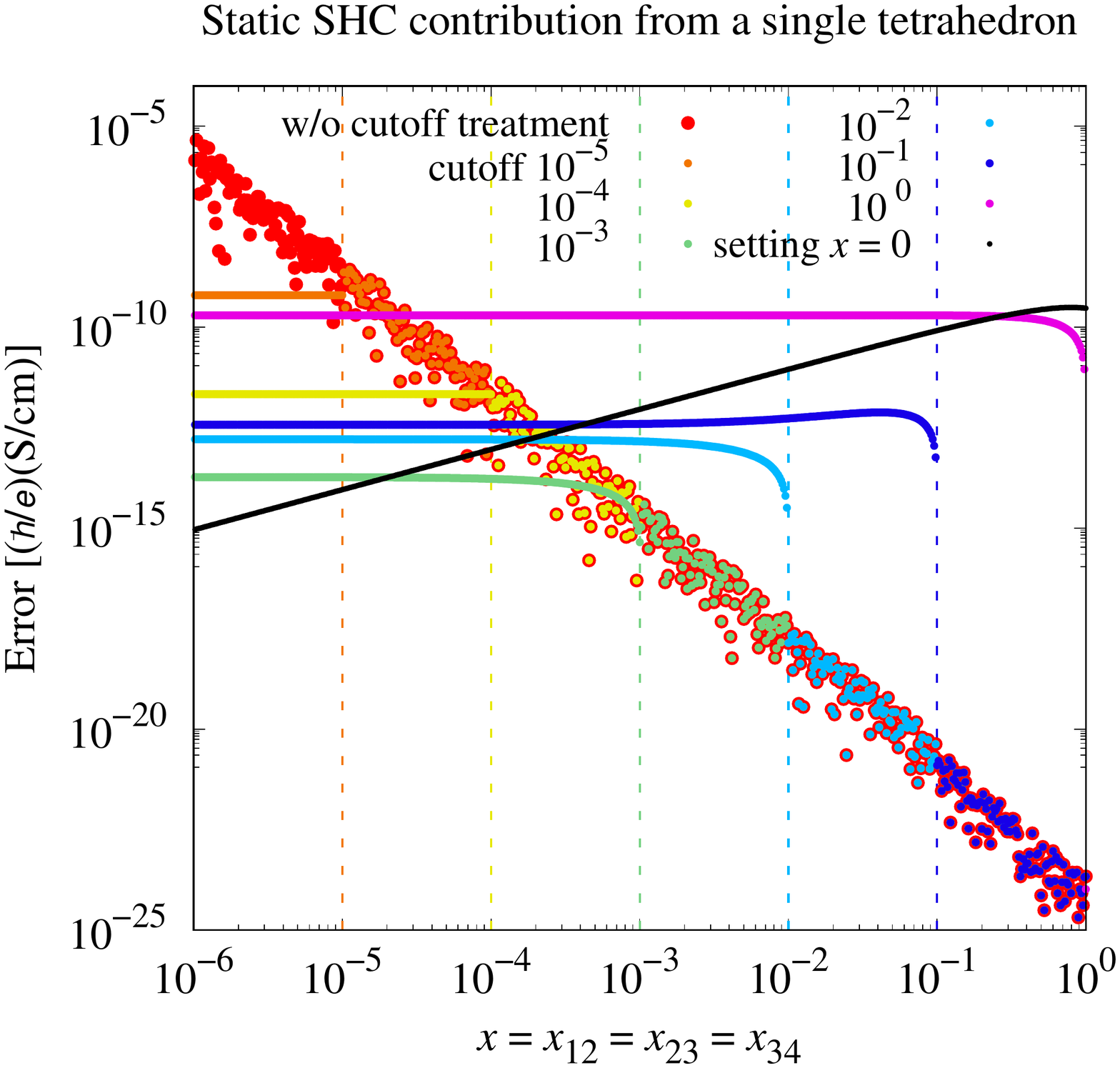}
    \caption{The difference (absolute value) between the value of static SHC from a single tetrahedron calculated using the cutoff treatment and the `exact' value versus $x=x_{12}=x_{23}=x_{34}$ [Eq.~\eqref{eq:x_ij}]. The `exact' value was obtained using quadruple precision numbers producing an ignorable round-off error within the shown $x$ range. The black curve indicates the absolute value of the difference between the exact value at a given $x$ and the value when $x$ is strictly zero.}
\label{fig:single_tetrahedron}
\end{figure}

Figure~\ref{fig:single_tetrahedron} illustrates how much the SHC values for a single tetrahedron with the cutoff treatment deviate from the `exact' value as a function of $x\,\, (=x_{12}=x_{23}=x_{34})$ [Eq.~\eqref{eq:x_ij}]. With the treatment, the integration values are fixed to be that of $|x| = \cutoff$; hence, the nearly horizontal lines emerge when $|x| < \cutoff$. Too small or too large $\cutoff$ leads to sizable errors, and $\cutoff = 10^{-3}$ displays the smallest error. For example, $\cutoff = 10^{-5}$ is too small, since the round-off error is not yet corrected by the treatment until this error gets very large. On the other hand, $\cutoff =10^{0}$ is too large because the energy eigenvalues at the four vertices are changed too much by the treatment. These facts are consistent with the results shown in Figs.~\ref{fig:convergence2}(a) and~\ref{fig:convergence2}(c). Without the treatment, the error increases with decreasing $x$ due to the round-off term $O(\epsilon)$. The slope of the diagonally scattered data in Fig.~\ref{fig:single_tetrahedron} is $-3$, from which we can infer that the round-off error is proportional to $1/x^3$. If all $|x_{ij}|$'s decrease at the same time, the error is not proportional to $1/x^2$ as Eqs.~\eqref{eq:xj_0_error} and \eqref{eq:xj_xk_0_error}. The reason for the slope being -3 is explained in Appendix \hyperlink{appendix:c}{C}.

Let $f(x)$ the formula evaluating the contribution to the integration from a single tetrahedron with a given $x$.  Then the slope of the black curve at small $x$ approaches $+1$ because $f(x) - f(0) = f'(0)x + O(x^2)$, where $f(0)$ is the limiting formula.  Also, $f'(0)$ can be found from the explicit formulas of the tetrahedron method, such as Eqs.~\eqref{eq:tetinteg1_2} and \eqref{eq:tetinteg3_2}.  Thus, the slope of the logarithmic plot is +1 when $x$ is small enough.

If we adopt a cutoff, $\cutoff$, the error follows the black curve when $x < \cutoff$. and the red scattered points when $x > \cutoff$.  The optimal cutoff is around $2 \cross 10^{-4}$ from the figure above since the error is minimized both when $x < \cutoff$ and $x > \cutoff$ at the same time.  Even if we adopt this cutoff, however, the error is comparable to or even larger than the error from our proposed simpler method with $\cutoff = 0.001$ (the green curve).

The Fermi-Dirac distribution we used is the step function, which cannot describe the smooth Fermi-Dirac distribution for finite temperatures. By scanning multiple Fermi levels and conducting a convolution of the integral of the non-dissipative part and the Fermi-Dirac distribution, we can take finite temperatures into account. This idea has been implemented in the WannierBerri code~\cite{tsirkin2021high}.

Our method can be applied to evaluate the integration of much more complicated functions with a different pole structure. From other correlation functions to Green's function or various self-energy, (for example, see Ref.~\cite{lihm2020phonon}) there are many quantities which can be expressed in the form of Eqs.~\eqref{eq:tetinteg1}--\eqref{eq:tetinteg3} after partial fraction decomposition.

The proposed tetrahedron method for the Kubo formula can also be straightforwardly extended to deal with the $\nk$-dependent, finite imaginary part of the quasiparticle self-energy.  The arguments of logarithms in Eqs.~\eqref{eq:tetinteg1_again} and \eqref{eq:tetinteg3_again} are now complex numbers instead of absolute values of real numbers.  For this extension, when calculating ${\rm log}A - {\rm log}B$, one should use the same branch cut for both log terms.  For example, if $A = e^{ia}$ and $B = e^{ib}$ with $a, b \in (-\pi, \pi]$, ${\rm log}A - {\rm log}B = i(a - b)$, while ${\rm log}(A/B) = i(a - b + 2n\pi)$ where $n$ is 0, +1, or -1 depending on $a$ and $b$.  Therefore, one should use ${\rm log}A - {\rm log}B$ and not ${\rm log}(A/B)$ in deriving the results such as Eqs.~\eqref{eq:tetinteg1_again} and \eqref{eq:tetinteg3_again}.

To summarize, we developed a new tetrahedron method to accurately evaluate the dissipative part of spectral functions. By discovering and solving a numerical problem due to rounding of floating-point arithmetic, we achieved improved convergence than the adaptive smearing method. A higher-order correction was also combined with our method. We demonstrated the advantage of our method by calculating the dynamical and static spin Hall conductivity of platinum. Our method can be applied to a wide range of other physical quantities. We will make the implementation of our proposed method publicly available through \wnt~\cite{pizzi2020wannier90} soon.

\section{ACKNOWLEDGEMENTS}
\begin{acknowledgments}
We thank Ji Hoon Ryoo for providing a code and directions to calculate SHC using MLWFs~\cite{ryoo2019computation}. We also thank Sophie Beck, Marco Gibertini, Jerome Jackson, Jason Kaye, Jae-Mo Lihm, and Jonathan R. Yates for useful discussions. This work was supported by the Institute for Basic Science (No. IBSR009-D1) and by the Creative-Pioneering Research Program through Seoul National University. Computational resources were provided by KISTI Supercomputing Center (Grant No. KSC-2020-INO-0078).
\end{acknowledgments}

\begin{figure}
    \includegraphics[trim={0cm 0cm 0cm 0cm}, width=0.53\columnwidth]{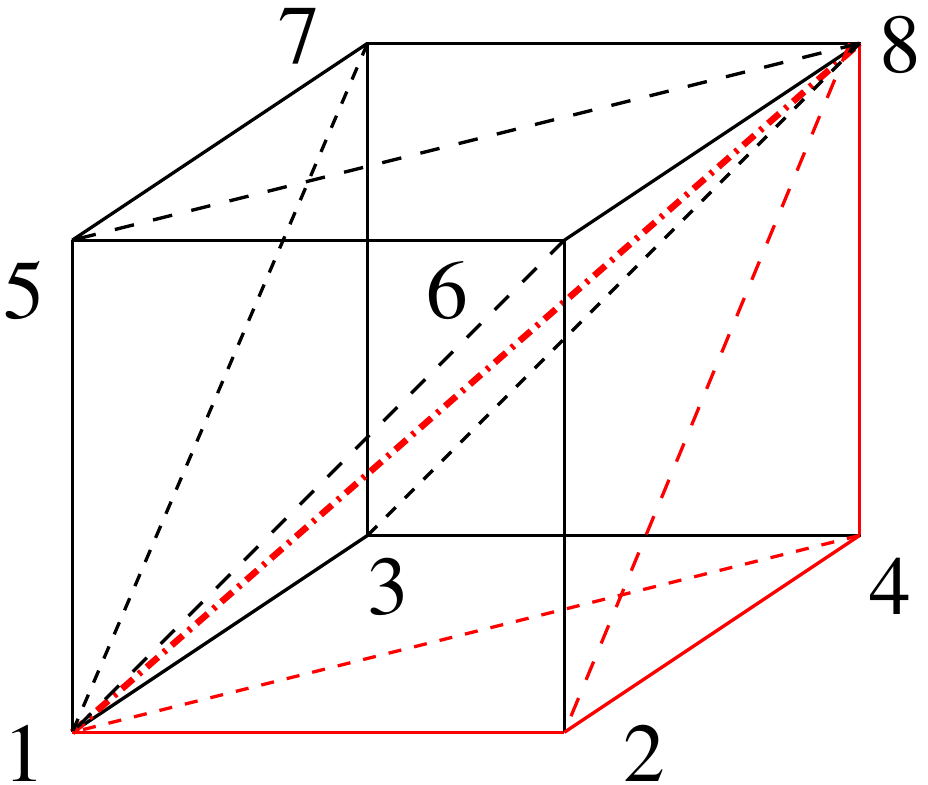}
    \caption{A schematic showing how six tetrahedra are generated from a given parallelepiped (cube in the figure): 1248 (red lines), 1628, 1568, 1758, 1378, and 1438. Here 18 is adopted to be the main diagonal (the red, dash-dotted line), and six edges (12, 13, 15, 84, 86, and 87) and five diagonals traversing a surface (dashed lines; 14, 17, 28, 38, 48) become edges of tetrahedra.}
    \label{fig:tetrahedron}
\end{figure}

\appendix
\section{STANDARD WAY TO CONSTRUCT TETRAHEDRA}
\renewcommand{\theequation}{A\arabic{equation}}
\setcounter{equation}{0}

\hypertarget{appendix:a}{The original way to construct tetrahedra is as follows~\cite{blochl1994improved}.} In the first Brillouin zone, $N$ $k$ points are chosen uniformly, and the $N$ parallelepipeds whose vertices are the neighboring $k$ points are built. Each parallelepiped is cut into six tetrahedra as depicted in Fig.~\ref{fig:tetrahedron}. As a result, the number of the simplexes becomes $6N$. The choice of the main diagonal ($18$ in the figure) is not unique, but it is advisable to select the shortest one.

When the Fermi surface or the surface $\hbar\omega-\Dnmk=0$ passes through a tetrahedron, a surface of constant energy $\epsilon_{\mk} = \eF$ or constant energy difference $\Dnmk = \hbar\omega$ cuts it into smaller polyhedra. Figures~\ref{fig:constsurface}(a)--\ref{fig:constsurface}(c) depict how the surface cuts a tetrahedron depending on the value of $\eF$ or $\hbar\omega$. Since $\epsilon_{\nk}$ and $\epsilon_{\mk}$ are linearly interpolated inside a single tetrahedron, each surface cuts a tetrahedron by a plane. The pieces cut by the surface are polyhedrons that can be constructed from smaller tetrahedra [Fig.~\ref{fig:constsurface}(d)].

\section{ANALYTIC TETRAHEDRON METHOD}
\renewcommand{\theequation}{B\arabic{equation}}
\setcounter{equation}{0}

\begin{figure*}
    \includegraphics[trim={0cm 1cm 0cm 1cm}, width=0.6\textwidth]{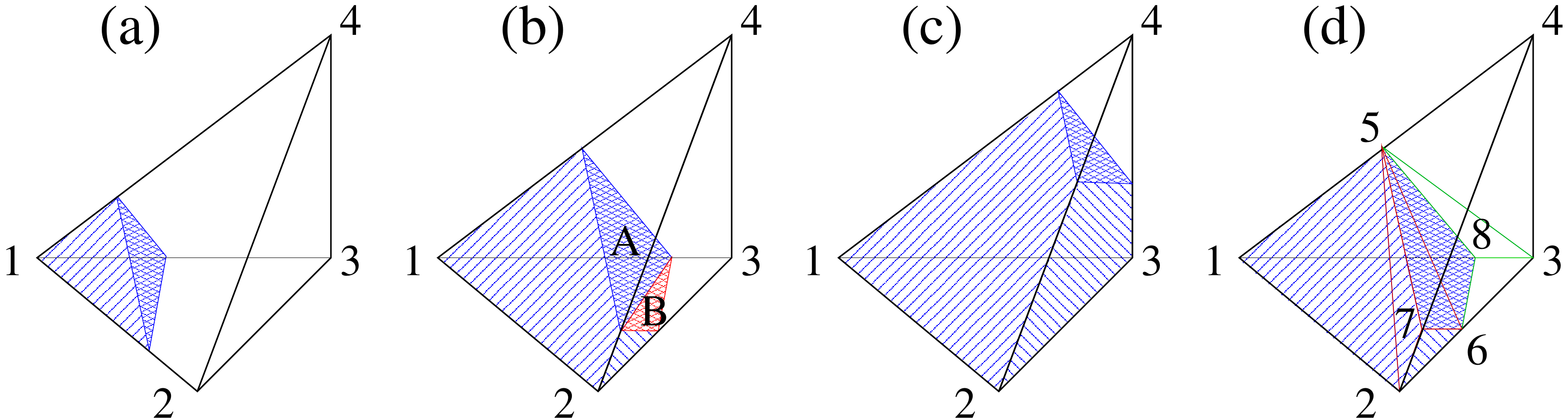}
    \caption{Illustrations of how the Fermi surface or the surface $\hbar\omega-\Dnmk=0$ cuts a tetrahedron in the Brillouin zone. It is assumed that the indices of the vertices are in order of increasing $\epsilon_{\mk}$ or $\Dnmk$, and they are denoted as $\epsilon_{i}$ or $\Delta_{i}$ for $i=1,2,3,4$. (a) $\epsilon_{1}<\eF<\epsilon_{2}$ or $\Delta_{1}<\hbar\omega<\Delta_{2}$. (b) $\epsilon_{2}<\eF<\epsilon_{3}$ or $\Delta_{2}<\hbar\omega<\Delta_{3}$; the tetragonal surface is cut into two triangular pieces, A and B. (c) $\epsilon_{3}<\eF<\epsilon_{4}$ or $\Delta_{3}<\hbar\omega<\Delta_{4}$. (d) A schematic showing that the polyhedron in panel (b) is constructed from three tetrahedra --- 1285 plus 5267 plus 5268.}
    \label{fig:constsurface}
\end{figure*}

\begin{table*}[]
    \centering
    \begin{tabularx}{0.95\textwidth}{c||l}
    \hline
    Range&Parametrization, $\partial(s_{1},s_{2},s_{3})/\partial(u,v)$\\
    \hline
    $\Delta_{1}<\hbar\omega<\Delta_{2}$&
    \scriptsize{$\begin{pmatrix} -y_{1} & -y_{1}\\ y_{2}&0\\0 & y_{3} \end{pmatrix}$}, 
    $y_{1}=\frac{\hbar\omega-\Delta_{1}}{\Delta_{2}-\Delta_{1}},  y_{2}=\frac{\hbar\omega-\Delta_{1}}{\Delta_{3}-\Delta_{1}}, y_{3}=\frac{\hbar\omega-\Delta_{1}}{\Delta_{4}-\Delta_{1}}$
    \\
    $\Delta_{2}<\hbar\omega<\Delta_{3}$&
    \scriptsize{$\begin{pmatrix} y_{1} & 0\\ 0 & y_{2}\\1-y_{1}-y_{3} & -y_{3} \end{pmatrix}$} for A, \scriptsize{$\begin{pmatrix} y_{1}-y_4 & -y_{4}\\ y_{4}-1 & y_{4}-1+y_{2}\\1-y_{1} & 0 \end{pmatrix}$} for B,
    $y_{1}=\frac{\hbar\omega-\Delta_{4}}{\Delta_{2}-\Delta_{4}},  y_{2}=\frac{\hbar\omega-\Delta_{1}}{\Delta_{3}-\Delta_{1}}, y_{3}=\frac{\hbar\omega-\Delta_{1}}{\Delta_{4}-\Delta_{1}}, y_{4}=\frac{\hbar\omega-\Delta_{3}}{\Delta_{2}-\Delta_{3}}$
    \\
    $\Delta_{3}<\hbar\omega<\Delta_{4}$&
    \scriptsize{$\begin{pmatrix} y_{1} & 0\\ 0 & y_{2}\\1-y_{1}-y_{3} & 1-y_{2}-y_{3} \end{pmatrix}$},
    $y_{1}=\frac{\hbar\omega-\Delta_{4}}{\Delta_{2}-\Delta_{4}},  y_{2}=\frac{\hbar\omega-\Delta_{4}}{\Delta_{3}-\Delta_{4}}, y_{3}=\frac{\hbar\omega-\Delta_{4}}{\Delta_{4}-\Delta_{1}}$
    \\
    \hline
    \end{tabularx}
    \caption{Parametrizations of the surface integrals [Eq.~\eqref{eq:tetinteg2}], assuming $\Delta_{1}\leq\Delta_{2}\leq\Delta_{3}\leq\Delta_{4}$.
    \label{tab:imaginary_part} For $\Delta_{2}<\hbar\omega<\Delta_{3}$, both triangles A and B shown in Fig.~\ref{fig:constsurface}(b) are parametrized.}
\end{table*}

\hypertarget{appendix:b}{Equations~\eqref{eq:tetinteg1} -- \eqref{eq:tetinteg3}} can be integrated over each tetrahedron using linear interpolations. Equations~\eqref{eq:tetinteg1} and~\eqref{eq:tetinteg3} are volume integrals, while Eq.~\eqref{eq:tetinteg2} is a surface integral. The integrands of the volume integrals are rational functions in $k$ space, so the results are associated with logarithmic functions. The volume integral is a triple integral of a linear function divided by another linear function
over a polyhedron, whose shape depends on $\epsilon_F$ [Fig.~\ref{fig:constsurface}]. The surface integral is a double integral on a plane cut by a surface with a constant energy difference $\Dnmk=\hbar\omega$ [Fig.~\ref{fig:constsurface}].

All the tetrahedra and triangles can be parametrized in new coordinate systems, $(s_1, s_2, s_3)$ and $(u,v)$, respectively, defined by
\begin{align}
\label{eq:k_ts}
    k_{i}=&\sum_{j}t_{ij}s_{j}\\
    \textbf{t}_{i}=&(t_{1i},t_{2i},t_{3i})\,,
\end{align}
and
\begin{equation}
\label{eq:param_uv}
    \begin{pmatrix} s_{1}\\s_{2}\\s_{3} \end{pmatrix}=\frac{\partial(s_{1},s_{2},s_{3})}{\partial(u,v)}\begin{pmatrix} u\\v \end{pmatrix},
\end{equation}
where $\textbf{t}_{i}$ is the vector from vertex 1 to vertex $(i+1)$ [Fig.~\ref{fig:constsurface}]. The Jacobian matrix in Eq.~\eqref{eq:param_uv} is given in Tab.~\ref{tab:imaginary_part}.
Under this change of variables, one vertex is set to be the origin, while the other two vertices are defined as $(u,v)=$(1,0), (0,1) for the triangular surface integral, and the other three vertices are defined as $(s_1,s_2,s_3)=$(1,0,0), (0,1,0), (0,0,1) for the tetrahedral volume integral.

The expressions for Eqs. \eqref{eq:tetinteg1} - \eqref{eq:tetinteg3} are:
\begin{align} 
\begin{split} \label{eq:tetinteg1_2}
    I_{1, nm}(\hbar\omega) ={}& -\frac{(1+x_1)(1+x_2)(1+x_3)}{6x_1^2 x_2^2 x_3^2 (x_1 - x_2)^2 (x_2 - x_3)^2 (x_3 - x_1)^2}\\
    & \times \frac{\textrm{det}(\textbf{t})}{\Delta_4 + \hbar\omega} \sum_{i = 1}^{4}F_{i}\left(\sum_{j=1}^{3}C_{ij}^{(1)}\xi_{j} + B_{i}^{(1)}\right)
\end{split}\\
\begin{split} \label{eq:tetinteg2_2}
        I_{2,nm}(\hbar\omega)={}&\frac{|\frac{\partial{\textbf{k}}}{\partial{u}}\cross\frac{\partial{\textbf{k}}}{\partial{v}}|}{|\nabla_{\textbf{k}}\Dnmk|}\cross\\
        &\int_{0}^{1}\int_{0}^{1-u} du\,dv\, F_{nm{\bf k}(u,v)}
\end{split}\\ 
\begin{split} \label{eq:tetinteg3_2}
    I_{3, nm} ={}&  \frac{(1+x_1)(1+x_2)(1+x_3)}{2x_1^2 x_2^2 x_3^2 (x_1 - x_2)^2 (x_2 - x_3)^2 (x_3 - x_1)^2}\\
    & \times \frac{\textrm{det}(\textbf{t})}{\Delta_4^2} \sum_{i = 1}^{4}F_{i}\left(\sum_{j=1}^{3}C_{ij}^{(3)}\xi_{j} + B_{i}^{(3)}\right)\,.
\end{split}
\end{align}
The expressions for $C_{ij}$, $B_{i}$, $|\nabla_{\textbf{k}}\Dnmk|$, and $F_{nm{\bf k}(u,v)}$ are given below. The definitions of the other used variables are presented in the main text.

For $a, b, c \in \{1, 2, 3\}$, define $b \equiv a + 1$ (mod 3), and $c \equiv a + 2$ (mod 3). In other words, they are cyclic.
\begin{align}
\label{eq:C_ij_1}
\begin{split}
    C_{aa}^{(1)} ={}&  -(x_b - x_c)^2 x_b^2 x_c^2\\
    &\times (3x_a^2 - 2(x_b + x_c)x_a + x_b x_c)\\
    C_{ba}^{(1)} ={}&  -(x_b - x_c)^2 x_b^2 x_c^2 x_a (1+x_b)(x_c - x_a)\\
    C_{ca}^{(1)} ={}& +(x_b - x_c)^2 x_b^2 x_c^2 x_a (1+x_c)(x_a - x_b)\\
    C_{4a}^{(1)} ={}&  -(x_b - x_c)^2 x_b^2 x_c^2(x_a - x_b)(x_c - x_a)\\
    B_{a}^{(1)} ={}&  x_a C_{4a}^{(1)}\\
    B_{4}^{(1)} ={}&  -x_1 x_2 x_3 (x_1 - x_2)^2 (x_2 - x_3)^2 (x_3 - x_1)^2
\end{split}
\end{align}
\begin{align}
\label{eq:C_ij_3}
\begin{split}
    C_{aa}^{(3)} ={}&  -(x_b - x_c)^2 x_b^2 x_c^2(1+x_a)\\
    &\times (2x_a^3 + (3-x_b-x_c)x_a^2 - 2(x_b + x_c)x_a + x_b x_c)\\
    C_{ba}^{(3)} ={}& (1+x_a)C_{ba}^{(1)}\\
    C_{ca}^{(3)} ={}& (1+x_a)C_{ca}^{(1)}\\
    C_{4a}^{(3)} ={}& (1+x_a)C_{4a}^{(1)}\\
    B_{a}^{(3)} ={}& x_a C_{4a}^{(3)}\\
    B_{4}^{(3)} ={}& B_{4}^{(1)}
\end{split}
\end{align}

The parametrizations for the surface integral $I_{2,nm}$ are written in terms of the values of energy differences. The detailed formulas are in Tab.~\ref{tab:imaginary_part}.
The gradient of $\Dnmk$ is constant on a single surface:
\begin{equation}
    |\nabla_{\textbf{k}}\Dnmk|=\sqrt{|\sum_{i,j,k}t^{-1}_{ik}t^{-1}_{jk}\Delta_{i}\Delta_{j}|}\,.
\end{equation}
Using the coefficients of $\Fnmk$ in the newly parametrized coordinate, $\Tilde{F_{0}},\Tilde{F_{1}},\Tilde{F_{2}}$,
\begin{align}
\begin{split}
    F_{nm{\bf k}(u,v)} =&\, F_1 + (F_2 - F_1)\,s_1(u,v) \\
     &+ (F_3 - F_1)\,s_2(u,v) + (F_4 - F_1)\,s_3(u,v)\\
    \equiv &\, \Tilde{F_{0}} + \Tilde{F_{1}}\,u + \Tilde{F_{2}}\,v\,,
\end{split}
\end{align}
we find that the surface integral is given by
\begin{equation}
    \int_{0}^{1}\int_{0}^{1-u} du\,dv\, F_{nm{\bf k}(u,v)}=\frac{\Tilde{F_{0}}}{2}+\frac{\Tilde{F_{1}}+\Tilde{F_{2}}}{6}\,.
\end{equation}
Also, $|\frac{\partial{\textbf{k}}}{\partial{u}}\cross\frac{\partial{\textbf{k}}}{\partial{v}}|$ in Eq. \eqref{eq:tetinteg2_2} is obtained using Eqs.~\eqref{eq:k_ts} --~\eqref{eq:param_uv} and the parametrization in Tab.~\ref{tab:imaginary_part} between the two coordinate systems $(s_1, s_2, s_3)$ and $(u,v)$. The integrals are hence calculable by collecting all the terms in Eqs.~\eqref{eq:tetinteg1_2} --~\eqref{eq:tetinteg3_2}.

\section{ANALYSIS OF THE ROUND-OFF ERROR SHOWN IN FIG.~\ref{fig:single_tetrahedron}}
\renewcommand{\theequation}{C\arabic{equation}}
\setcounter{equation}{0}

\hypertarget{appendix:c}{According} to Eqs.~\eqref{eq:xj_0_error} and \eqref{eq:xj_xk_0_error}, the round-off error is proportional to $O(\epsilon)/x^2$ if only one of $|x_j|$ or $|x_j - x_k|$ approaches~0. In Fig.~\ref{fig:single_tetrahedron}, however, the slope in logarithmic scale is $-3$, implying the error is proportional to $O(\epsilon)/x^3$.

In order to understand this behavior, suppose that $x_1$, $x_2$, $x_3$, $x_1 - x_2$, $x_2 - x_3$, and $x_3 - x_1$ are all proportional to $O(x)$. We are using log$(x) \doteq O(x) + O(\epsilon)$ or log1P$(x) \doteq O(x) + O(x\epsilon)$, so  $s=0$ or $1$ for $\xi_l = \textrm{log} (1+x_l) \doteq O(x_l) + O(x_l^{s} \epsilon)$, respectively.

\begin{align}
\label{eq:x_all_0_error}
\begin{split}
    &\frac{\sum_l C_{il}\xi_{l}+ B_i}{x_1^2 x_2^2 x_3^2 (x_1 - x_2)^2 (x_2 - x_3)^2 (x_3 - x_1)^2}\qquad\quad\\
    &\doteq \frac{O(1)}{x^{12}} [C_{ii}(O(x_i)+O(x_i^s\epsilon)) + C_{ij}(O(x_j)+O(x_j^s\epsilon)) \\
    &\qquad\qquad +
    C_{ik}(O(x_k)+O(x_k^s\epsilon)) + B_i ] \\
    &= \frac{1}{x^{12}} [O(x^8)(O(x)+O(x^s\epsilon)) + O(x^8)(O(x)+O(x^s\epsilon)) \\
    &\qquad\qquad +
    O(x^8)(O(x)+O(x^s\epsilon)) + O(x^9) ] \\
    &= \frac{1}{x^{12}}[O(x^{12}) + O(x^8)O(x^s\epsilon)]\\
    &= \frac{O(x^{4-s}) + O(\epsilon)}{x^{4-s}}\,.
\end{split}
\end{align}

\newpage

In the second to last equality in Eq.~\eqref{eq:x_all_0_error}, we have used the fact that the coefficients of $O(x^{9})$, $O(x^{10})$, and $O(x^{11})$ should vanish in order to have a non-divergent value in the limit $x\to0$ if $\epsilon=0$. Hence, the round-off error term is
\begin{equation}
    \frac{O(\epsilon)}{x^{4-s}}\,.
\end{equation}
The error is proportional to $1/x^3$ as shown in Fig.~\ref{fig:single_tetrahedron} since log1P was used in our implementation, i.\,e.\,, $s=1$.

\bibliography{main}

\end{document}